# Disentangling Brain Graphs: A Note on the Conflation of Network and Connectivity Analyses

*Sean L. Simpson*[1,2*], *Paul J. Laurienti*[2,3]

[1]*Biostatistical Sciences, Wake Forest School of Medicine, Winston-Salem, NC 27157*

[2]*Laboratory for Complex Brain Networks, Wake Forest School of Medicine, Winston-Salem, NC 27157*

[3]*Radiology, Wake Forest School of Medicine, Winston-Salem, NC 27157*

*Correspondence:

Sean L. Simpson, Ph.D.

Department of Biostatistical Sciences,

Wake Forest School of Medicine

Medical Center Blvd.,

Winston-Salem, NC, 27157, USA

slsimpso@wakehealth.edu

ABSTRACT

Understanding the human brain remains the Holy Grail in biomedical science, and arguably in all of the sciences. Our brains represent the most complex systems in the world (and some contend the universe) comprising nearly one hundred billion neurons with septillions of possible connections between them. The structure of these connections engenders an efficient hierarchical system capable of consciousness, as well as complex thoughts, feelings, and behaviors. Brain connectivity and network analyses have exploded over the last decade due to their potential in helping us understand both normal and abnormal brain function. Functional connectivity (FC) analysis examines functional associations between time series pairs in specified brain voxels or regions. Brain network analysis serves as a distinct subfield of connectivity analysis in which associations are quantified for all time series pairs to create an interconnected representation of the brain (a brain network), which allows studying its systemic properties. While connectivity analyses underlie network analyses, the subtle distinction between the two research areas has generally been overlooked in the literature, with them often being referred to synonymously. However, developing more useful analytic methods and allowing for more precise biological interpretations requires distinguishing these two complementary domains.

**Introduction**

Brain connectivity and network analyses have exploded over the last decade, moving to the forefront of the neuroimaging field. Their importance in our understanding normal and abnormal brain function has been well documented (Biswal et al., 2010; Sporns, 2010). Functional connectivity (FC) analysis examines functional associations between time series pairs in specified brain voxels or regions (Biswal 1995). Functional brain network analysis serves as a distinct subfield of FC analysis in which associations are quantified for all time series pairs to create an interconnected representation of the brain (a brain network). The resulting connection matrix is often thresholded to create a binary adjacency matrix that retains "significant" connections (edges) while removing weaker ones, but weighted (continuous) network analyses are gaining traction due to recent methodological advances (Rubinov and Sporns, 2011). The appeal of the network approach is that it allows studying how systemic properties of the brain relate to behavioral and health outcomes (Bullmore and Sporns, 2009; Telesford et al., 2011;

Simpson et al., 2013; Bassett and Bullmore, 2009). Here we focus on *functional* connectivity and network analyses, but the commentary in this note applies to structural analyses as well.

As we have noted in Simpson and Laurienti (2015) and elsewhere, the systemic organization present in brain networks confers much of our brains' functional abilities as connections may be lost due to an adverse health condition, but compensatory connections may develop as a result in order to maintain organizational consistency and functional performance as illustrated in Figure 1. Thus, different groups (or individuals) may exhibit differences in connectivity while retaining the same network structure. In reality, the brain likely only partially compensates for damaged connections as has been discussed in both the brain network science literature (Fornito et al., 2015; Qi et al., 2010; Fischer et al., 2014) and neuroscience literature more generally (Barulli and Stern, 2013), especially in adult brains. Hence, connectivity and network analyses may provide distinct, but complementary insight into individual and group differences, making joint or hybrid analyses crucial to our understanding of normal and abnormal brain function. In the following sections we briefly delineate methods for connectivity and network analyses and discuss the importance of joint and hybrid methodology for expanding the scope of neuroscience research.

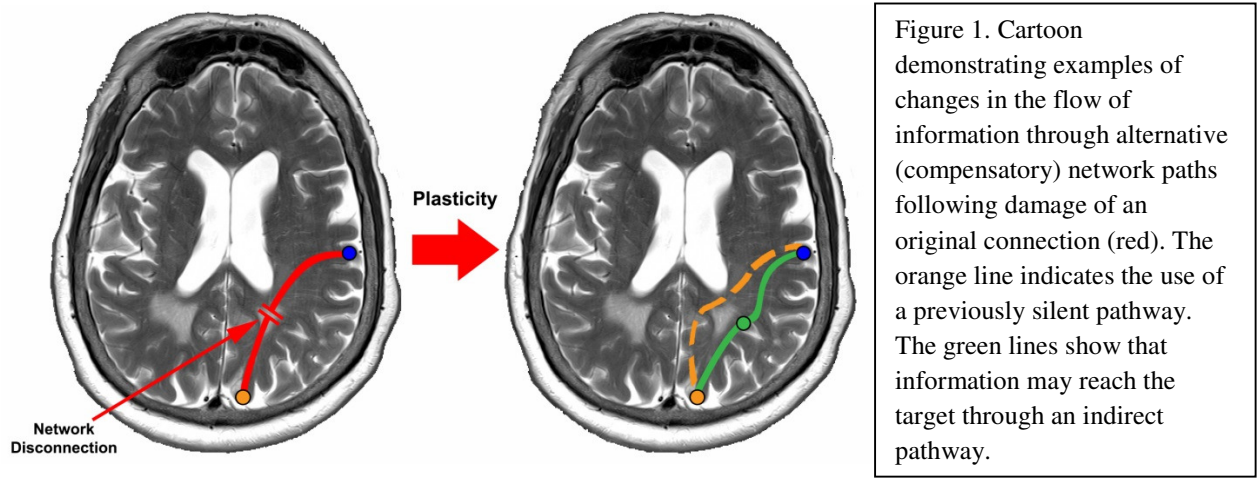

Figure 1. Cartoon demonstrating examples of changes in the flow of information through alternative (compensatory) network paths following damage of an original connection (red). The orange line indicates the use of a previously silent pathway. The green lines show that information may reach the target through an indirect pathway.

**Connectivity Methods**

Functional connectivity methods comprise both methods for estimating the functional association between time series pairs in specified brain regions and methods for drawing inference from these estimated connections as a function of covariates of interest (e.g., disease status). Estimation methods fall into three categories: association measures, modeling approaches, and partitioning methods. Pairwise correlation is the most commonly used association measure, with measures such as coherence, mutual information, and generalized synchronization employed less frequently. Partial correlation provides a multivariate analog of pairwise correlation that better distinguishes direct from indirect connections, but presents computational challenges which have been the focus of ongoing research (Chen et al., 2013). Modeling approaches for estimating connectivity are diverse yet remain relatively limited in use due to the acceptance of more easily implementable association measures. These modeling methods, surveyed in Simpson et al. (2013) and Bowman et al. (2015), often inherently allow identifying group-related connectivity differences which remains a subsequent step when association methods are employed. Partitioning methods, which group brain areas together in sets that exhibit more within set functional similarity than between set similarity, include independent component analysis (ICA) and cluster analysis approaches (e.g., K-means clustering, fuzzy clustering, hierarchical clustering).

Most inferential approaches for identifying difference in functional connectivity either stem from the modeling-based estimation methods noted above or rely on mass-univariate comparisons between the employed association measure (often correlation) of the connections with a multiple testing correction applied. Under this mass-univariate umbrella, Smith et al. (2013) treated the partial correlation of each edge as a covariate in a general linear (regression) model (GLM) predicting various participant phenotypes (e.g., behavioral measure). Further details on connectivity methods can be found in Simpson et al. (2013) and Bowman et al. (2015).

**Network Methods**

Network methods aim to describe, model, or draw inference from fully constructed networks (derived from the estimated connectivity patterns). Descriptive methods aim to quantify systemic properties such as clustering ("local communication"), path length ("global communication"), modularity, order $l$ degree distribution (Bagrow et al., 2008), etc. As with connectivity methods, most inferential network methods, which aim to identify differences in systemic properties, rely on univariate approaches. Network metrics (e.g., clustering, path length) at the network or nodal level are often rudimentarily compared employing a t-test or ANOVA like techniques. More sophisticated univariate approaches include the network based statistic (NBS) and spatial pairwise clustering (SPC) (Zalesky et al., 2012). Both methods are predicated on connection by connection comparisons and then subsequently aggregate the results of these comparisons to identify clusters of edge-based differences. A related multivariate approach, partial least squares (PLS), identifies functional connectivity patterns (i.e., edge combinations) that optimally covary with experimental design parameters such as group status or task condition (Wold, 1985; McIntosh et al., 1996; Berman et al., 2014; Mišić et al., 2014; Shen et al., 2015). While often labeled as network methods, one could categorize these three approaches as connectivity methods given their focus on sets of connections and the designation of network methods as those that emphasize systemic properties of connectivity patterns. While these approaches have led to important insights, gaining a deeper understanding of normal and abnormal changes in complex functional organization demands methods that leverage the wealth of data present in an entire brain network. As noted in the Introduction, this systemic organization confers much of our brains' functional abilities as functional connections may be lost due to an adverse health condition, but compensatory connections may develop as a result in order to maintain organizational consistency and functional performance. Thus, we believe that gaining insight into this organization requires a multivariate modeling framework that allows assessing the effects of systemic properties (network measures) and phenotype (e.g., demographics, disease status, etc.) on the overall network structure. That is, if we have

$$\text{Data} \begin{cases} \boldsymbol{Y}_i: \text{network of participant } i \\ \boldsymbol{X}_i: \text{covariate information} \end{cases},$$

we wish to accurately estimate the probability density function of the network given the covariates $P(\boldsymbol{Y}_i|\boldsymbol{X}_i, \boldsymbol{\theta}_i)$, where $\boldsymbol{\theta}_i$ are the parameters that relate the covariates to the network structure. We have made strides in developing such a framework both with exponential random graph models (ERGMs) (Simpson et al., 2011, 2012) and mixed models (Simpson and Laurienti, 2015), but more work is needed on refining these approaches, and developing new ones.

The ERGM and mixed modeling frameworks provide complementary multivariate approaches for analyzing the brain at the network level, that is, for assessing systemic infrastructural properties of the entire network as opposed to just properties of specific nodes or connections. ERGMs allow efficiently representing network data by modeling its global structure as a function of local substructural properties. However, they are limited in their ability to examine specific connections, compare groups of networks, and assess the relationship between networks and phenotypic characteristics. Mixed models generally allow examining specific connections, are well-suited for group comparisons, and enable assessing the relationship between networks and phenotypic traits, complementing ERGMs, but are limited in their ability to capture the inherent complex dependence structure of brain networks. Our approach in Simpson and Laurienti (2015) attempts to adapt mixed models to the brain network context and account for this dependence structure. It also serves as what could be considered a rudimentary connectivity/network analysis hybrid method given its use of dyads as outcome variables while accounting for dependence and network properties via the random effects and network metric fixed effects parameters. Given their flexibility, mixed models may provide the machinery necessary to develop the needed hybrid methods for furthering our understanding of brain function. At a minimum, they will be beneficial in joint network/connectivity analyses in conjunction with an appropriate connectivity method.

**Conclusion**

As with all biological systems, studying the brain at various levels (micro, meso, macro) remains paramount, especially given the hierarchical nature of its physiology. In our context this requires analyzing both connectivity properties (specific interregional connections) and higher level network properties (systemic architecture). An alternate conception puts these two sets of properties under the same network analysis umbrella, as opposed to viewing them as distinct interrelated domains, with connections representing the basic level and graph properties representing the systemic, higher level. Both conceptions necessitate a multi-level approach, which is particularly important given the ability of the brain to compensate at the network level for "damage" to specific connections. Thus, drawing more precise biological conclusions and advancing our understanding of brain function demand hybrid analyses that derive insight both at the individual connection and network level. These analyses may result from jointly assessing connectivity and network properties with separate methodologies, akin to multi-modal neuroimaging analysis, or from novel hybrid methodologies. Moving a new field forward necessitates refining the language and concepts used within it. Properly distinguishing connectivity from network analyses provides a step in this direction and allows better leveraging the complementary information contained in these two domains.

**Acknowledgements**

This work was supported by NIBIB K25 EB012236-01A1 (Simpson), and Wake Forest Older Americans Independence Center (P30 21332) and the Sticht Center on Aging (Laurienti).